\documentclass[aps,preprint]{revtex4}
\usepackage{epsfig}

\def\bea{\begin{eqnarray}}
\def\eea{\end{eqnarray}}
\def\bec{\begin{center}}
\def\ec{\end{center}}

\def\beq{\begin{equation}}
\def\eeq{\end{equation}}

\begin{document}
\draft
\tighten
\preprint{KAIST-TH 2005/18}
\title{\large \bf Moduli stabilization and the pattern of soft SUSY breaking terms}
\author{
Kiwoon Choi\footnote{kchoi@hep.kaist.ac.kr}}
\address{
Department of Physics, Korea Advanced Institute of Science
and Technology, Daejeon 305-701, Korea}
\date{\today}
%\maketitle
%%%%%%%%%%%%%%%%%%%%%%%%%%%%%%%%%%%%%%%%%%%%%%%%%%%%%%%

\vspace{1cm}

\begin{abstract}
In string compactification preserving $N=1$ SUSY, moduli fields
(including the string dilaton) are plausible candidates for the
messenger of SUSY breaking at low energy scales. In a scenario
that moduli-mediated SUSY breaking is significant, the pattern of
soft SUSY breaking terms depends crucially  on how the light
moduli with mass $m \lesssim {\cal O}(8\pi^2 m_{3/2})$ are
stabilized. We discuss the correspondence between the pattern of
soft terms and the stabilization mechanism of light moduli
within the framework of  4D effective supergravity  which
is generalized to include a SUSY-breaking uplifting potential
which might be necessary to get the
phenomenologically viable de-Sitter (or Minkowski) vacuum.
% is achieved by  as in the recent
%KKLT construction of de-Sitter  vacua in Type IIB string theory.
In some special case, light moduli can be stabilized by controllably small
perturbative corrections to the K\"ahler potential, yielding
%$m_X\lesssim m_{3/2}$, the moduli $F$-component $F_X/X \sim m_{3/2}$, and
the soft terms dominated by the moduli-mediated contribution. In
more generic situation, light moduli are stabilized by
non-perturbative effects encoded in the superpotential and a quite
different pattern of soft terms emerges:
%$m_X\sim 8\pi^2 m_{3/2}$ and $F_X/X\sim m_{3/2}/8\pi^2$, thus
the anomaly-mediated soft terms become comparable to
the moduli-mediated ones. Such mixed moduli-anomaly mediated soft terms
can be described by a mirage messgenger scale hierarchically lower than
$M_{\rm Planck}$, and lead to low energy superparticle masses
qualitatively different from those of other mediation models
such as mSUGRA scenario, gauge-mediation, and anomaly-mediation.
%A possibility to generate  flavor-violating soft terms
%which would have nontrivial implications for $B$ physics
%is briefly discussed also.
%%%%%%%%%%%%%%%%%%%%%%%%%%%%%%%%%%%%%%%%%%%%%%%%%%%%%%%
%%%%%%%%%%%%%%%%%%%%%%%%%%%%%%%%%%%%%%%%%%%%%%%%%%%%%%%%
\end{abstract}
%%%%%%%%%%%%%%%%%%%%%%%%%%%%%%%%%%%%%%%%%%%%%%%%%%%%%%%%
\pacs{}
\maketitle

\section{introduction}

Low energy supersymmetry (SUSY) is one of the prime candidates for
physics beyond the standard model at TeV scale
\cite{Nilles:1983ge}. One of the key questions on low energy SUSY
is the origin of soft SUSY breaking terms of the visible
gauge/matter superfields in the low energy effective lagrangian
\cite{kane}. Most of the phenomenological aspects of
low energy SUSY are determined by those soft terms which
are presumed to be induced by the auxiliary components of some messenger
fields. In string theory, moduli fields including the string
dilaton are plausible candidates for the messenger of SUSY
breaking. In addition to  string moduli, the 4-dimensional
supergravity (SUGRA) multiplet provides a model-independent source
of SUSY breaking, i.e. the anomaly mediation
\cite{Randall:1998uk}, which induces a soft mass
 $m_{\rm soft}\sim m_{3/2}/8\pi^2$.
%where $m_{3/2}$ is the gravitino mass.
To identify the dominant source of soft terms, one needs to
compute the relative ratios between different auxiliary components
including the auxiliary component of the 4D SUGRA multiplet, which
requires an understanding of how the messenger moduli are
stabilized at a nearly 4D Poincare invariant vacuum. If some
moduli $X$ are stabilized with a heavy mass $m_X\gg 8\pi^2
m_{3/2}$, their auxiliary components are typically suppressed as
$F_X/X \sim m_{3/2}^2/m_X\ll m_{3/2}/8\pi^2$.
As a result, the
soft terms mediated by such heavy moduli are negligible compared
to the anomaly-mediated ones. On the other hand, light moduli
$T$ with $m_T\lesssim 8\pi^2m_{3/2}$ generically have $F_T/T \gtrsim
m_{3/2}/8\pi^2$, and then the soft terms mediated by light moduli
can be comparable to or dominate over the anomaly-mediated
contributions. As we will see, different values of  the
anomaly to moduli-mediation ratio lead to {\it qualitatively
distinguishable} patterns of low energy superparticle spectrum.
This means that one can probe the mechanism of moduli
stabilization through the low energy superparticle masses which
might be observable at future collider experiments.

In this talk, I discuss the correspondence between the soft terms
and the stabilization mechanism of light moduli for string
compactifications which can realize the low energy SUSY at TeV
scale together with the high scale gauge coupling unification at
$M_{GUT}\sim 2\times 10^{16}$ GeV \cite{choi1}. Theoretical framework of the
discussion is the 4D effective SUGRA
%of light moduli and
%visible gauge/matter fields with which one can compute
which allows the computation of moduli $F$-components and
the resulting soft terms in a controllable approximation scheme. In
order to obtain a phenomenologically viable de-Sitter (or
Minkowski) vacuum, the effective SUGRA  is generalized to include
a SUSY-breaking uplifting potential as in the recent KKLT
construction of de-Sitter  vacua in Type IIB string theory \cite{kklt}. In
section II, I briefly discuss 4D effective SUGRA which contains an
uplifting sector. I then present in section III several different
examples of light moduli stabilization,
%and compute the resulting anomaly to moduli-mediation ratio.
including  the case that the light moduli are stabilized by
controllably small perturbative corrections to the K\"ahler
potential \cite{hebecker} as well as the KKLT stabilization
\cite{kklt} achieved by
non-perturbative superpotential \cite{gauginocondensation}. The perturbative K\"ahler
stabilization is possible in a rather limited situation and leads
to the soft terms dominated by moduli-mediation, while the
stabilization by non-perturbative superpotential can be
implemented in more generic situation and leads to the soft terms
receiving comparable contributions from both the moduli mediation
and the anomaly mediation \cite{choi1}.
% comparable to the moduli-mediation.
Section IV is devoted to the discussion of the low energy
superparticle spectrum for generic values of the anomaly to
moduli-mediation ratio \cite{choi2,yama,adam}.
%and also a brief discussion of how the
%SUSY flavor/CP problems can be naturally avoided in the mixed
%moduli-anomaly mediated SUSY breaking scenario.

\section{4D effective SUGRA with uplifting sector}

Quite often, moduli stabilization with low energy SUSY
yields hierarchical moduli masses: some moduli $X$ get large
masses $m_X\gg 8\pi^2 m_{3/2}$, while the other moduli $T$ remain
to be light with $m_T\lesssim 8\pi^2 m_{3/2}$. For instance, the
moduli  stabilized by quantized fluxes \cite{gkp} are typically heavy if
the compactification scale $M_{\rm com}$ and the string scale
$M_{\rm string}$ are hierarchically higher than $m_{3/2}$,
%which is presumed to be close to the TeV scale:
%2\times 10^{16}$ GeV, while $m_{3/2}$ is hierarchically lower than $M_{GUT}$:
e.g. $m_X\sim M_{\rm com}^n/M_{\rm string}^{n-1}\gg 8\pi^2
m_{3/2}$ for the moduli masses induced by $n$-form flux. The heavy
moduli masses are dominated by the SUSY-preserving part: $m_X
\,\simeq\, \left\langle e^{K/2}\partial_X^2
W/\partial_X\partial_{X^*}K\right\rangle
%\,\gg\, 8\pi^2 \langle e^{K/2}W\rangle
$ where $K$ and $W$ are the K\"ahler potential and the superpotential
of 4D SUGRA, and then the resulting auxiliary components are
suppressed as $F^X/X\,=\,\left\langle
e^{K/2}D_{X^*}W^*/X\partial_X\partial_{X^*}K \right\rangle
\,\sim\, m_{3/2}^2/m_X$ \cite{choi1}. Such heavy moduli are decoupled from low
energy SUSY breaking, thus can be safely integrated out, leaving
an effective 4D SUGRA which describes the stabilization of light
moduli. Our theoretical framework is such an effective SUGRA of
light moduli which contains also an uplifting sector which might
be necessary to get a phenomenologically viable de-Sitter (or
Minkowski) vacuum \cite{kklt}. Here we will be focusing on the
compactifications with $M_{\rm com}\gtrsim M_{GUT}=2\times 10^{16}$
GeV which can accomodate the
successful 4D gauge coupling unification at $M_{GUT}$.
%in which
%the uplifting sector originates from anti-brane.
The uplifting sector might originate from
a SUSY-breaking anti-brane in the underlying string
compactification.
Although it appears to break SUSY explicitly,
the uplifting sector can be consistently accomodated in 4D $N=1$ SUGRA
through a Goldstino operator
in $N=1$ superspace \cite{choi1}.

The effective action of light moduli $T_i$ and the visible gauge
and matter superfields, $W^a_\alpha$ and $Q^I$, can be written as
(in the unit with the 4D Planck scale $M_{Pl}=1$):
%\footnote{It has been noted that when the uplifting potential
%originates
%from anti-brane whose tension is appropriately red-shifted by
%small warp factor, which might be necessary
%to make the total vacuum energy density to be small enough,
%its low energy
%consequence can be described  by a $D$-type spurion operator $\theta^2\bar{\theta}^2{\cal P}_{\rm lift}$
%alone since the potentialy possible $F$-type spurion operator $\theta^2\Gamma$ gives
%negligible effects which are suppressed by additional powers of
%small warp factor.}
\begin{eqnarray}
\label{N=1}
S_{\rm eff}&=&\int d^4x \sqrt{g^C} \,\left[\,
\int d^4\theta \,\left\{
CC^*\Big(\,-3\exp(-K/3)\,\Big) -C^2C^{2*}\theta^2\bar{\theta}^2 {\cal P}_{\rm lift}
\,\right\}\right.\nonumber \\
&&+\,\left.\left\{
\int d^2\theta
\left(\frac{1}{4}f_a W^{a\alpha}W^a_\alpha
+C^3W\right)
+{\rm h.c.}\right\}\,\right],
\end{eqnarray}
where
\bea
K&=&K_0(T_i,T_i^*)+Z_I(T_i,T_i^*)Q^IQ^{I*},
\nonumber \\
{\cal P}_{\rm lift}&=&{\cal
P}_0(T_i,T_i^*)+X_I(T_i,T_i^*)Q^IQ^{I*},
\nonumber \\
W\,=\,&W_0(T_i)&+\,\frac{1}{6}\lambda_{IJK}Q^IQ^JQ^K,
\qquad
f_a\,=\, f_a(T_i).
\eea
Here we are using the superconformal formulation of 4D SUGRA
with chiral compensator superfield $C$,
and $g^C_{\mu\nu}$ is the 4D metric
in superconformal frame which is related to the Einstein frame metric
$g^E_{\mu\nu}$ as
$g^C_{\mu\nu}=(CC^*)^{-1}e^{K/3}g^E_{\mu\nu}$.
%Note that one needs an off-shell formulation of $N=1$ SUGRA
%in order to describe the coupling between the SUSY-breaking
%anti-brane (or a sector in which $N=1$ SUSY is non-linearly realized)
%and the other sector in which $N=1$ SUSY
%is linearly realized.
For simplicity, we choose the superconformal gauge in which both
the fermionic component of $C$ and the scalar auxiliary component
of SUGRA multiplet are vanishing, and then ignore the dependence
of SUGRA multiplets other than the spacetime metric-dependence.
There still remains a residual super Weyl invariance under the
transformation \bea C\rightarrow e^{-2\tau}C, \quad
g^C_{\mu\nu}\rightarrow e^{2(\tau+\tau^*)}g^C_{\mu\nu}, \quad
\theta^\alpha \rightarrow e^{-\tau+2\tau^*}\theta^\alpha, \eea
where $\tau$ is a complex constant, and the uplifting spurion
operator $\theta^2\bar{\theta}^2{\cal P}_{\rm lift}$
 should be invariant under this super Weyl transformation
in order to keep the consistency of formulation. For the
consistency with the full $N=1$ local SUSY, one might replace the
Grassmann coordinate $\theta^\alpha$ in the spurion operator  by
the Goldstino superfield $\Lambda^\alpha =\theta^\alpha+$
Goldstino-terms \cite{wess}:
\bea
\theta^2\bar{\theta}^2{\cal P}_{\rm lift}\quad\longrightarrow
\quad\Lambda^2\bar{\Lambda}^2{\cal P}_{\rm lift},
\eea which will not affect our
subsequent discussion.

The $D$-type  uplifting spurion operator in (\ref{N=1}) does not affect the
standard on-shell relations for the SUSY breaking auxiliary
components (in the Einstein frame):
\begin{eqnarray}
\label{approx-F} \frac{F^C}{C}\,=
\,\frac{1}{3}\partial_iK_0F^i+m_{3/2}^*, \qquad F^i\,=
\,-e^{K_0/2}\left(\partial_i\partial_{\bar{j}}K_0\right)^{-1}\left(D_jW_0\right)^*,
\eea where $F^C$ and $F^i$ denote the auxiliary $F$-components of
$C$ and $T_i$, respectively, and  $m_{3/2}=  e^{K_0/2}W_0$. On the
other hand, the moduli potential  is modified to include the
uplifting potential from spurion operator:
%which is crucial for $\langle V_0\rangle=0$:
\begin{eqnarray}
\label{approx-potential}
V_{TOT}&=& V_F+V_{\rm lift},
\eea
where
$$
V_F=
\left(\partial_i\partial_{\bar{j}}K_0\right)F^iF^{j*}-3e^{K_0}|W_0|^2
%\left(\left(\partial_i\partial_{\bar{j}}K_0\right)^{-1}D_iW_0(D_jW_0)^*-3|W_0|^2\right)
$$
is the standard  $F$-term potential of the $N=1$ SUGRA, and the
uplifting potential is given by
$$
V_{\rm lift}=
e^{2K_0/3}{\cal P}_0(T_i,T_i^*).
$$

It should be stressed that this uplifting potential is fundamentally different
from the $D$-term potential of the $N=1$ SUGRA which is associated with
4D gauge symmetry.
To see the difference explicitly, let us consider a 4D SUGRA with
an anomalous $U(1)$ gauge symmetry (but without the uplifting sector) \cite{bkq} under which
\bea
\delta T=i\delta_{GS},\quad
\delta\Phi^I =iq_I\Phi^I,\eea
where the non-linear transformation of $T$  ($\delta_{GS}=$ real constant)
is introduced to realize the Green-Schwarz (GS) anomaly cancellation mechanism,
and $q_I$ is the $U(1)$ charge of generic chiral superfield $\Phi^I$
other than $T$.
%In addition to the GS anomaly cancellation
%through the $T$-dependent $U(1)_X$ gauge kinetic function $f$,
%the non-linear transformation of $T$ induces a $T$-dependent Fayet-Iliopoulos (FI)
%term $\xi_{FI}=\delta_{GS}\partial_TK$.
The resulting $N=1$ SUGRA potential is given by
\bea
V_{N=1}=V_F+V_D=\Big[\,
\left(\partial_A\partial_{\bar{B}}K\right)F^AF^{B*}-3e^K|W|^2\,\Big]+\Big[\,
\frac{1}{2}g^2D^2\,\Big]
\eea
where $F^A$ is the $F$-component of $\Phi^A=(T,\Phi^I)$,
$g$ is the $U(1)$ gauge coupling,
and the auxiliary $D$-component of the $U(1)$ vector multiplet contains the
$T$-dependent Fayet-Iliopoulos (FI) term $\xi_{FI}=\delta_{GS}\partial_TK$ arising from the non-linear
$U(1)$ transformation of $T$:
\bea
\label{d1}
D=-i\delta\Phi^A\partial_AK=
\xi_{FI}+q_I\Phi^I\partial_IK=
i\frac{\left(\partial_A\partial_{\bar{B}}K\right)\delta\Phi^A F^{B*}}{m_{3/2}}.
\eea
Here the last expression of $D$ is  based on the $U(1)$ invariance of
$W$ and is an identity which is valid as long as $W\neq 0$.
The above expression of $D$ already shows a fundamental difference between $V_D=g^2D^2/2$
and $V_{\rm lift}$. The $D$-term potential
$V_D$ can never shift the supersymmetric AdS minimum
($\langle F^A\rangle=0$ and $\langle W\rangle\neq 0$) of $V_F$
since $\langle V_D\rangle_{\rm SUSY-AdS}=0$ is a global minimum
of $V_D$ \cite{choi1}.
On the other hand,  $V_{\rm lift}$ generically shifts
the supersymmetric AdS minimum of $V_F$ to a SUSY-breaking true vacuum
of $V_{TOT}=V_F+V_{\rm lift}$.
%the $U(1)$ gauge invariance enforces
%that the FI-term $\xi_{FI}=\delta_{GS}\partial_TK$ in $D$ is
%precisely cancelled by the  other $D$-term contribution
%$q_I\Phi^I\partial_IK$
%for any field configuration with
%$F^A=0$ and $W\neq 0$.

Even for a SUSY-breaking solution with $F^A\neq 0$, the possible
value of $V_D$ is severely constrained.
Applying the stationary condition
\bea
\langle\partial_AV_{N=1}\rangle=0
\eea
together with the $U(1)$ invariance of $K$ and $W$, one finds
the following relation on the VEVs of the $F$ and $D$-components
\cite{kawamura}:
\bea
\label{constraint}
i\Big(\,\frac{1}{M_{Pl}^2}K_{A\bar{B}}F^AF^{B*}
-m_{3/2}^2+M_V^2\,\Big)D\,=\,
\frac{g^2}{2}\delta\Phi^A\left[\partial_A\ln({\rm Re}(f))\right]D^2
\nonumber \\
\,+\,F^AF^{B*}\Big[K_{C\bar{B}}\partial_A\delta\Phi^C
+\delta\Phi^C\partial_CK_{A\bar{B}}\Big],
\qquad\qquad\eea
where $f$ is the $U(1)$
gauge kinetic function, and
\bea
\label{gaugeboson}
M_V^2=g^2K_{A\bar{B}}\delta\Phi^A\delta\Phi^{B*}=
g^2M_{Pl}^2\delta_{GS}^2\partial_T\partial_{T^*}K+...
\eea
denotes the $U(1)$ gauge boson mass.
This shows that $M_V^2\gg m_{3/2}M_{Pl}$
as long as $T$ is stabilized at a value which  does {\it not}
give a hierarchically small  $\langle
\partial_T\partial_{T^*}K\rangle$.
%, which might be required to get
%the high compactification scale $M_{\rm com}\sim
%M_{GUT}$.
Then the  constraint (\ref{constraint}) implies
$\langle D\rangle\ll m_{3/2}M_{Pl}$ since
$\langle F^A\rangle={\cal O}(m_{3/2}M_{Pl})$ or smaller, finally leading to
$\langle V_D\rangle\ll m_{3/2}^2M_{Pl}^2$.
In fact, if the modulus $T$ corresponds to the string dilaton or
the volume modulus, the corresponding
$M_V^2$ is rather close to $M_{Pl}^2$
for the range of $\langle T\rangle$ which can be compatible with
the high compactification scale $M_{\rm com}\gtrsim M_{GUT}$.
Then  $\langle V_D\rangle
={\cal O}\,(m_{3/2}^4)$
which is too small to be useful for uplifting the negative vacuum energy density
$-3m_{3/2}^2M_{Pl}^2$ in $V_{N=1}$.
%This difficulty to obtain $\langle V_D\rangle ={\cal O}(m_{3/2}^2M_{Pl}^2)$
%while avoiding a hierarchically small vacuum value of
%the moduli K\"ahler metric  originates essentially from
%the gauge invariance of the theory.
On the other hand, there is {\it no} constraint on $V_{\rm lift}$
such as (\ref{constraint}) since
there is no associated gauge symmetry.
As a result, one can freely adjust the
size of $\langle V_{\rm lift}\rangle$ to be ${\cal O}\,(m_{3/2}^2M_{Pl}^2)$
in order to get the desired nearly flat de-Sitter vacuum.

 Once $K_0$, $W_0$ and ${\cal P}_0$ are given, one can compute
$F^C$ and $F^i$  by minimizing $V_{TOT}=V_F+V_{\rm lift}$ under the fine tuning
condition for $\langle V_{TOT}\rangle =0$. The resulting soft  terms
of canonically normalized visible fields can be written as
\begin{eqnarray}
{\cal L}_{soft}&=&-\frac{1}{2}M_a\lambda^a\lambda^a-\frac{1}{2}m_I^2|\phi^I|^2
-\frac{1}{6}A_{IJK}y_{IJK}\phi^I\phi^J\phi^K+{\rm h.c.},
\end{eqnarray}
where $\lambda^a$ are gauginos, $\phi^I$ are the scalar components
of the matter superfields $Q^I$, and
$y_{IJK}=\lambda_{IJK}/\sqrt{e^{-K_0}Z_IZ_JZ_K}$ are the
canonically normalized Yukawa couplings. Since the
anomaly-mediated contributions to soft masses can be comparable to
the moduli-mediated contributions, i.e. $F^C/8\pi^2C\sim F^i/T_i$,
we need to include both contributions in a consistent manner,
yielding the following form of soft parameters at energy scales
just below $M_{GUT}$ \cite{choi1}:
\begin{eqnarray}
\label{soft1}
M_a&=& F^i\partial_i\ln\left({\rm Re}(f_a)\right) +\frac{b_ag_a^2}{8\pi^2}\frac{F^C}{C},
\nonumber \\
A_{IJK}&=&
-F^i\partial_i\ln\left(\frac{\lambda_{IJK}}{e^{-K_0}Z_IZ_JZ_K}\right)-
\frac{1}{16\pi^2}(\gamma_I+\gamma_J+\gamma_K)\frac{F^C}{C},
\nonumber \\
m_I^2&=& \frac{2}{3}(V_F+V_{\rm
lift})+e^{2K_0/3}Z_I^{-1}X_I-F^iF^{\bar{j}}\partial_i\partial_{\bar{j}}\ln
\left(e^{-K_0/3}Z_I\right)
\nonumber \\
&-&\frac{1}{32\pi^2}\frac{d\gamma_I}{d\ln\mu}\left|\frac{F^C}{C}\right|^2
+ \frac{1}{16\pi^2}\left\{ (\partial_{i}{\gamma}_I)
F^i\left(\frac{F^C}{C_0}\right)^* +{\rm h.c.}\right\},
\end{eqnarray}
where $b_a$ and $\gamma_I$ are the one-loop beta function
coefficients and the anomalous dimension of $Q^I$ which are given
by
%$\frac{dg_a}{d\ln \mu}=\frac{b_a}{8\pi^2} g_a^3,
%\frac{d\ln Z_I}{d\ln \mu}=\frac{1}{8\pi^2}\gamma_I$:
\bea b_a&=&-\frac{3}{2}{\rm tr}\left(T_a^2({\rm
Adj})\right)+\frac{1}{2}\sum_i {\rm tr}\left(T^2_a(Q^I)\right)
\nonumber \\
\gamma_I&=&2C_2(Q^I)-\frac{1}{2}\sum_{JK}|y_{IJK}|^2 \quad
(\,\sum_a g_a^2T_a^2(Q^I)\equiv C_2(Q^I)\bf{1}\,)
\nonumber \\
\partial_{i}\gamma_I
&=&-\frac{1}{2}\sum_{JK}|y_{IJK}|^2\partial_i\ln\left(\frac{\lambda_{IJK}}{e^{-K_0}Z_IZ_JZ_K}\right)
-2C_2(Q^I)\partial_i\ln\left({\rm Re}(f_a)\right), \nonumber \eea
where $\omega_{IJ}=\sum_{KL}y_{IKL}y^*_{JKL}$ is assumed to be
nearly diagonal.

Note that the uplifting operator
$\Lambda^2\bar{\Lambda}^2{\cal P}_{\rm lift}=
\Lambda^2\bar{\Lambda}^2[{\cal P}_0+X_IQ^IQ^{I*}]$ affects the soft
scalar mass as \bea \Delta m_I^2 =
e^{2K_0/3}\left[\frac{2}{3}{\cal P}_0+Z_I^{-1}X_I\right]
=\frac{2}{3}\frac{V_{\rm lift}}{M_{Pl}^2}+e^{2K_0/3}Z_I^{-1}X_I, \eea where the
first term is the consequence of lifting the vacuum energy
density by $V_{\rm lift}$,
while the second term originates from
the matter-Goldstino contact term
$X_I\Lambda^2\bar{\Lambda^2}Q^IQ^{I*}$ in the uplifting operator.
%(Note that generically
%the soft scalar mass is  affected
%by any kind of contribution to the vacauum energy density.)
% which should be taken into account in the moduli
%stabilization in which the uplifting sector plays a non-negligible
%role for SUSY-breaking.
In the KKLT-type moduli stabilization, one needs $V_{\rm lift}=
{\cal O}(m_{3/2}^2M_{Pl}^2)$ in order to get a phenomenologically viable
de-Sitter (or Minkowski) vacuum. Still $X_I$ can be negligibly
small compared to $m_{3/2}^2$ if the uplifting sector is
sequestered from the visible sector. In the KKLT compactification
of Type IIB string theory, the uplifting operator originates from
anti-$D3$ brane ($\bar{D}_3$) which is stabilized at the end of
an warped throat \cite{kklt}. On the other hand, if one wishes to accommodate the
gauge coupling unification at $M_{GUT}\sim 2\times 10^{16}$ GeV,
the visible sector should be assumed to live on $D$-branes located
at the {\it un-warped region} of the Calabi-Yau space. As the
$D$-branes of $Q^I$  are separated from the $\bar{D}_3$ of
$\Lambda^\alpha$ by a long throat, the coefficient $X_I$ of the
superspace contact interaction between $Q^I$ and
$\Lambda^\alpha$ is expected to be highly suppressed. Unfortunately, a
reliable calculation of $X_I$ requires a detailed knowledge of the
underlying string compactification, which is not available at this
moment.
%In the discussion of the low energy phenomenology of
%(\ref{soft1}), we will simply assume that
%the uplifting sector is well sequestered from the visible sector, thus
%$X_I$ are small enough to be ignored.
%Note that $M_a$ and $A_{IJK}$ are independent of the sequestering
%assumption not aff
%phenomenological aspects of the

\section{Some examples}

\subsection{Stabilization by perturbative K\"ahler corrections}

If the light moduli have a {\it flat} potential when the leading
order form of the K\"ahler potential is used, some of the light moduli might
be stabilized by
% the competition between different
(controllably small) perturbative corrections to the K\"ahler potential.
 A simple example is the stabilization of the volume
modulus $T={\cal V}_{CY}^{2/3}+ic_4$ in Type IIB string compactification
 on a Calabi-Yau (CY) orientifold with positive Euler number
\cite{hebecker}.
(Here ${\cal V}_{CY}$ is the volume of the CY orientifold,
and $c_4$ is a zero mode of the RR 4-form field.)
 After the dilaton
and complex structure moduli get heavy masses $m\sim M_{\rm
com}^3/M_{\rm st}^2\gg 8\pi^2 m_{3/2}$ by 3-form fluxes,
%and thus are integrated out,
the 4D effective SUGRA of the  K\"ahler moduli takes the no-scale form
at leading order in $\alpha^\prime$ and string loop expansions,
thus gives a flat potential of K\"ahler moduli.
Considering only the overall volume modulus
while  including the higher order corrections to the K\"ahler potential,
one finds \cite{hebecker}
\bea
K_0&=&-3\ln(T+T^*)+\frac{\xi_1}{(T+T^*)^{3/2}}-\frac{\xi_2}{(T+T^*)^2},
\nonumber \\
W_0&=&\omega_0,
\qquad
V_{\rm lift}\,=\,\frac{D_0}{(T+T^*)^{2}}, \eea
where $\xi_1$
is the coefficient of $\alpha^\prime$
correction which is positive for a positive Euler number,
$\xi_2$  is the coefficient of string
loop correction,
$\omega_0\sim m_{3/2}$ (in the unit with
$M_{Pl}=1$) is a constant induced by 3-form fluxes, and $V_{\rm lift}$ is
the uplifting potential generated by anti-$D3$ brane.
%Although $\ell=2$ for $V_{\rm lift}$ originating from $\bar{D}_3$,
%here we will take
%$\ell$ as a constant of order unity.
% which should
%be fine-tuned to make the final vacuum energy density small enough.
It is then straightforward to see that, if $\xi_{1,2}>0$,
 ${\rm Re}(T)$ is
stabilized by the competition between the $\alpha^\prime$ and
string loop corrections at a vacuum value
\bea
\langle {\rm Re}(T)\rangle\simeq
5.1\xi_2^2/\xi_1^2,
\eea
for which
 \bea
\frac{F^T}{(T+T^*)}\,\simeq\, m_{3/2},\quad
 \frac{1}{8\pi^2}\frac{F^C}{C}\,\simeq\, 6.3\times 10^{-3}
\frac{\xi_1}{(T+T^*)^{3/2}}m_{3/2},
\nonumber \\
m_T\,\simeq\, 0.8 \frac{\xi_1^{1/2}}{(T+T^*)^{3/4}}m_{3/2}, \quad
V_{\rm lift}\,\simeq \, 0.12\frac{\xi_1}{(T+T^*)^{3/2}}m_{3/2}^2M_{Pl}^2. \eea

In order for the above volume modulus stabilization to be a reliable approximation,
one needs the K\"ahler corrections
to be significantly smaller than the leading order term:
\bea
\left\langle\frac{\xi_1}{(T+T^*)^{3/2}}\right\rangle
\,\simeq\,
%2\sqrt{2}\xi_2/(T+T^*)^2\simeq
3.1\times 10^{-2}\frac{\xi_1^4}{\xi_2^3}\,\ll\, 1,
\eea
Then the anomaly to modulus mediation ratio in this K\"ahler stabilization
scheme is negligibly small:
\bea
\left(\frac{1}{8\pi^2}\frac{F^C}{C}\right)\left(\frac{F^T}{T+T^*}\right)^{-1} \,\ll\, 6.3\times 10^{-3},
\eea
and the  soft terms are dominated by the modulus-mediated contributions of
${\cal O}(m_{3/2})$. Note that the size of the uplifting
potential which is required to get a nearly flat dS vacuum
is not significant compared to $m_{3/2}^2M_{Pl}^2$,
thereby the effects of the uplifting sector on soft terms can be ignored also.
 An interesting feature of this K\"ahler stabilization  is that
the RR axion ${\rm Im}(T)$ can be identified as the QCD axion solving the strong CP problem,
though it might suffer from  the cosmological difficulty since its decay
constant $f_a\sim M_{GUT}$ \cite{choi4}.

One difficulty of the above K\"ahler stabilization
of the volume
modulus is that it works only for
$\xi_1>0$ which requires
the Euler number $\chi=2(h_{1,1}-h_{1,2})>0$.
In order to get $\omega_0$ hierarchically lower than $M_{Pl}$,
one might need many independent NS and RR 3-form fluxes, i.e.
need a large value of $h_{1,2}$.
In such case, the number of  K\"ahler moduli
is also large, $h_{1,1}> h_{1,2}$, thus there remains many
unstabilized K\"ahler moduli.
To avoid this difficulty, one might
stabilize some of the K\"ahler moduli  by non-perturbative
superpotential. However then the K\"ahler moduli
unfixed by non-perturbative superpotential
do not have a flat potential, thereby
can not be stabilized by a controllably small
K\"ahler correction. Another potential difficulty of
the K\"ahler stabilization is that it
generically predicts $m_T\lesssim m_{\rm soft}$, thus might
suffer from the cosmological moduli problem when the soft masses
of visible fields have the weak scale size.

\subsection{KKLT-type stabilization by non-perturbative superpotential}

A more interesting possibility is to stabilize the light moduli by
non-perturbative superpotential. The simplest example would be the
KKLT stabilization described by the following effective SUGRA
 \bea
\label{kklt1}
K_0\,=\,-n_0\ln(T+T^*),
\quad
W_0\,=\,\omega_0-Ae^{-aT},
\quad
V_{\rm lift}\,=\,\frac{D_0}{(T+T^*)^{\ell}}, \eea
where $A$ is a constant of order unity (in the unit with $M_{Pl}=1$).
Stabilization of
the Calabi-Yau volume modulus in Type IIB can be described by this
form of effective SUGRA with $n_0=3$ and $\ell=2$ after (i) the heavy
dilaton and complex structure moduli are integrated out, and (ii) the
hidden gaugino condensation and anti-$D3$ brane are introduced to generate
$Ae^{-aT}$  and  $V_{\rm lift}$ \cite{kklt}, respectively. Stabilization of
the heterotic dilaton might be described also by this form of effective
SUGRA with $n_0=1$ after (i) the K\"ahler and complex structure moduli
get heavy masses by the  geometric tortion and NS fluxes \cite{lukas},
and (ii)  the hidden gaugino condensation and an
anti-NS5 brane are  introduced. At any rate, one easily finds that
$T$ is stabilized at
\bea
a\langle T\rangle\simeq
 \ln(A/\omega_0)\simeq \ln(M_{Pl}/m_{3/2})
\eea
for which
\bea
\frac{F^T}{(T+T^*)}\,\simeq\,
%\frac{3}{2aT}\frac{\partial_T\ln(V_{\rm lift})}{\partial_TK_0}m_{3/2}\simeq
\frac{3\ell}{2n_0}\frac{m_{3/2}}{\ln(M_{Pl}/m_{3/2})},\quad
\frac{1}{8\pi^2}\frac{F^C}{C}\,\simeq\, \frac{m_{3/2}}{8\pi^2},
\nonumber \\
m_T\,\simeq\,  m_{3/2}\ln(M_{Pl}/m_{3/2}), \quad
 V_{\rm lift}\,\simeq\, 3m_{3/2}^2M_{Pl}^2.
 \eea

If the  low energy SUSY is realized within this KKLT stabilization scheme,
one would have $\ln(M_{Pl}/m_{3/2})\sim 4\pi^2$, and then
the corresponding anomaly to modulus mediation ratio is essentially of order
unity:
\bea
\left(\frac{1}{8\pi^2}\frac{F^C}{C}\right)\left(\frac{F^T}{T+T^*}\right)^{-1}
\,\simeq\, \frac{n_0}{3\ell}\frac{\ln(M_{Pl}/m_{3/2})}{4\pi^2}\,=\,
{\cal O}(1).
\eea
As will be discussed in the next section, such mixed modulus-anomaly mediation
can give a highly distinctive pattern of superparticle masses at TeV scale.
An interesting feature of the KKLT stabilization is the little hierarchy
structure:
\bea
m_T\sim m_{3/2}\ln(M_{Pl}/m_{3/2})\sim
m_{\rm soft}\left[\ln(M_{Pl}/m_{3/2})\right]^2,
\eea
for which the cosmological moduli and gravitino problems
can be avoided even when the soft masses have the weak scale size.

Unlike the K\"ahler stabilization, the
KKLT stabilization  can be easily
generalized to the multi K\"ahler moduli case. For each K\"ahler
modulus $T_i$, there can be a stack of $D7$ branes warpping the
corresponding 4-cycle. Then the gaugino condensation on those $D7$
branes will generate the non-perturbative
superpotential $A_ie^{-a_iT_i}$  stabilizing $T_i$.
As an explicit example, let us consider a Calabi-Yau compactification
with two K\"ahler moduli whose effective SUGRA is given by \cite{quevedo}:
\bea
\label{kklt2}
K_0&=& -2\ln\left[(T_1+T^*_1)^{3/2}-(T_2+T^*_2)^{3/2}\right]
\nonumber \\
W_0&=& \omega_0-A_1e^{-a_1T_1}-A_2e^{-a_2T_2},
\nonumber \\
V_{\rm lift}&=&\frac{D_0}{\left[(T_1+T^*_1)^{3/2}-(T_2+T^*_2)^{3/2}\right]^{2\ell/3}},
\eea
where $A_{1,2}$ are constants of order unity (in the unit with $M_{Pl}=1$),
We then find
\bea
a_i\langle T_i\rangle\simeq
 \ln(A_i/\omega_0)\simeq \ln(M_{Pl}/m_{3/2}) \quad
(i=1,2),
\eea
and
\bea
\frac{F^1}{(T_1+T_1^*)}\,\simeq\,
\frac{F^2}{(T_2+T_2^*)}\,\simeq\,\frac{3\ell}{2n_0}\frac{m_{3/2}}{\ln(M_{Pl}/m_{3/2})},\qquad
\qquad
\nonumber \\
\frac{1}{8\pi^2}\frac{F^C}{C}\,\simeq\, \frac{m_{3/2}}{8\pi^2},\quad
m_{T_{1,2}}\,\simeq\,  m_{3/2}\ln(M_{Pl}/m_{3/2}) ,
\quad V_{\rm lift}\,\simeq\, 3m_{3/2}^2M_{Pl}^2,
 \eea
which are essentially same as the case of single K\"ahler modulus.

If the light moduli are stabilized by a race-track form of superpotential,
the anomaly to moduli mediation ratio can be significantly enhanced,
leading to the anomaly-dominated SUSY breaking.
For instance, for the effective SUGRA
 \bea
K_0=-3\ln(T+T^*),\quad
%\nonumber \\
W_0=A_1e^{-a_1T}-A_2e^{-a_2T},\quad
%\nonumber\\
V_{\rm lift}=\frac{D_0}{(T+T^*)^{\ell}}
\eea
with $A_{1,2}={\cal O}(1)$
and the race-track fine-tuning $|\,a_1-a_2|\simeq
(a_1+a_2)/\ln(M_{Pl}/m_{3/2})$,
one finds \cite{choi1}
\bea
\frac{F^T}{(T+T^*)}\,\simeq\,
%\frac{3}{2aT}\frac{\partial_T\ln(V_{\rm lift})}{\partial_TK_0}m_{3/2}\simeq
\frac{3\ell}{4}\frac{m_{3/2}}{[\ln(M_{Pl}/m_{3/2})]^2},\quad
\frac{1}{8\pi^2}\frac{F^C}{C}\,\simeq\, \frac{m_{3/2}}{8\pi^2},
\nonumber \\
m_T\,\simeq\,  \frac{4}{3}m_{3/2}[\ln(M_{Pl}/m_{3/2})]^2, \quad
 V_{\rm lift}\,\simeq\, 3m_{3/2}^2M_{Pl}^2,
\nonumber \\
\left(\frac{1}{8\pi^2}\frac{F^C}{C}\right)\left(\frac{F^T}{T+T^*}\right)^{-1}
\,\simeq\, \frac{2}{3\ell}\frac{[\ln(M_{Pl}/m_{3/2})]^2}{4\pi^2}
\,=\,{\cal O}(4\pi^2).
\eea

\section{Low energy phenomenology of the mirage mediation}

Let us discuss how the low energy superparticle spectrum behaves as
the anomaly to moduli mediation ratio varies. The soft terms of
(\ref{soft1}) renormalized at the scale just below $M_{GUT}$ can
be written as \bea M_a&=& M_0\left[\,
1+\frac{\ln(M_{Pl}/m_{3/2})}{8\pi^2} b_ag^2_{GUT}\alpha\,\right],
\nonumber \\
A_{IJK}&=&M_0\left[\,a_{IJK}
-\frac{\ln(M_{Pl}/m_{3/2})}{16\pi^2}(\gamma_I+\gamma_J+\gamma_K)\alpha\,\right]
\nonumber \\
m_I^2&=& |M_0|^2\left[\,
c_I-\frac{\ln(M_{Pl}/m_{3/2})}{4\pi^2}\Big(C_2(Q^I)-
\frac{1}{4}a_{IJK}|y_{IJK}|^2\Big)\alpha\nonumber \right. \\
&& \qquad -\,
\left.\frac{[\ln(M_{Pl}/m_{3/2})]^2}{32\pi^2}\frac{d\gamma_I}{d\ln\mu}\alpha^2\,
\right]
\eea
where
\bea
M_0&\equiv & F^i\partial_i\ln\left({\rm Re}(f_a)\right),
\nonumber \\
a_{IJK}M_0&\equiv&
-F^i\partial_i\ln\left(\frac{\lambda_{IJK}}{e^{-K_0}Z_IZ_JZ_K}\right),
\nonumber \\
c_I|M_0|^2&\equiv & -F^iF^{\bar{j}}\partial_i\partial_{\bar{j}}\ln
\left(e^{-K_0/3}Z_I\right)+e^{2K_0/3}Z_I^{-1}X_I \eea represent the
moduli-mediated soft terms (including the soft scalar mass from
the matter-Goldstino contact term), and
\bea \alpha\equiv
\left(\frac{4\pi^2}{\ln(M_{Pl}/m_{3/2})}\right) \left(\frac{1}{M_0}\right)
\left(\frac{1}{4\pi^2}\frac{F^C}{C}\right) \eea
parameterizes the {\it anomaly to moduli mediation ratio}.
Note that $\alpha$ is of order unity when the two mediations
give comparable contributions to the gaugino masses.

Taking into account 1-loop RG evolution, the low energy gaugino
masses at the renormalization point $\mu$ are given by \bea
\label{lowgaugino} M_a(\mu)&=&M_0\left[\,
1-\frac{1}{4\pi^2}b_ag_a^2(\mu)\ln\left(\frac{M_{GUT}}{(M_{Pl}/m_{3/2})^{\alpha/2}\mu}\right)
\,\right], \eea where $g_a(\mu)$ are the running gauge couplings at
scale $\mu$. As for the low energy values of $A_{IJK}$ and
$m_I^2$, if $y_{IJK}\lesssim 1/\sqrt{8\pi^2}$ {\it or}
$a_{IJK}=c_I+c_J+c_K=1$ for the $I$-$J$-$K$ combination with
$y_{IJK}\gtrsim 1/\sqrt{8\pi^2}$, they are given by \bea
\label{lowsoft} && A_{IJK}(\mu)\,= \,M_0\left[\,
a_{IJK}+\frac{1}{8\pi^2}(\gamma_I(\mu)
+\gamma_J(\mu)+\gamma_K(\mu))
\ln\left(\frac{M_{GUT}}{(M_{Pl}/m_{3/2})^{\alpha/2}\mu}\right)\,\right],
\nonumber \\
&& m_I^2(\mu)\,= \,
|M_0|^2\left[\,c_I-\frac{1}{8\pi^2}Y_I\Big(\sum_Jc_JY_J\Big)
g_Y^2(\mu)\ln\left(\frac{M_{GUT}}{\mu}\right)\right.
\nonumber \\
&&\,\,\,\left. +\,\,\frac{1}{4\pi^2}
\left\{\gamma_I(\mu)-\frac{1}{2}\frac{d\gamma_I(\mu)}{d\ln\mu}
\ln\left(\frac{M_{GUT}}{(M_{Pl}/m_{3/2})^{\alpha/2}\mu}\right)\right\}
\ln\left(\frac{M_{GUT}}{(M_{Pl}/m_{3/2})^{\alpha/2}\mu}\right)\,\right]
, \eea where $\gamma_I(\mu)$ denote the running anomalous
dimensions at $\mu$ and $Y_I$ is the $U(1)_Y$ hypercharge of
$Q^I$.

The above results of low energy soft masses show an
interesting feature:
%the anomaly-mediated contributions at
%$M_{GUT}$, i.e. the $\alpha$-dependent part, cancel the RG
%evolution of the modulus-mediated soft masses from $M_{GUT}$ to
% $(m_{3/2}/M_{Pl})^{\alpha/2}M_{GUT}$ other than the evolution
% of scalar mass due to a large Yukawa coupling or a nonzero
% $\sum_Ic_IY_I$.
the low energy gaugino and (1st and 2nd generation) scalar masses
in the mixed modulus-anomaly mediation
with the messenger scale $M_{GUT}$
are same as those in the pure
moduli-mediation with a {\it mirage messenger scale} \bea
M_{\rm mirage}\equiv (m_{3/2}/M_{Pl})^{\alpha/2}M_{GUT}. \eea
As an immediate consequence of this  mirage mediation,
when $\alpha$ increases from $\alpha=0$ (pure moduli-mediation)
to $\alpha\simeq 2$ (mixed moduli-anomaly mediation), the superparticle masses
at TeV scale get closer to each other as the mass splitting due to the RG running
(over the scales between $M_{GUT}$ and $M_{\rm mirage}$) is
cancelled.  Note that the mirage messenger scale does not correspond to a physical
threshold scale. Still the physical gauge coupling unification
scale is $M_{GUT}$, and the Kaluza-Klein and string threshold
scales are a little above $M_{GUT}$.
% $aT\approx \ln(M_{Pl}/m_{3/2})\approx \ln(M_{GUT}/\mbox{TeV})$,
%We get a particularly interesting gaugino masses at $M_{SUSY}\sim 1$ TeV

\begin{figure}[t]
\begin{center}
\begin{minipage}{16cm}
\centerline{
{\hspace*{-.2cm}\psfig{figure=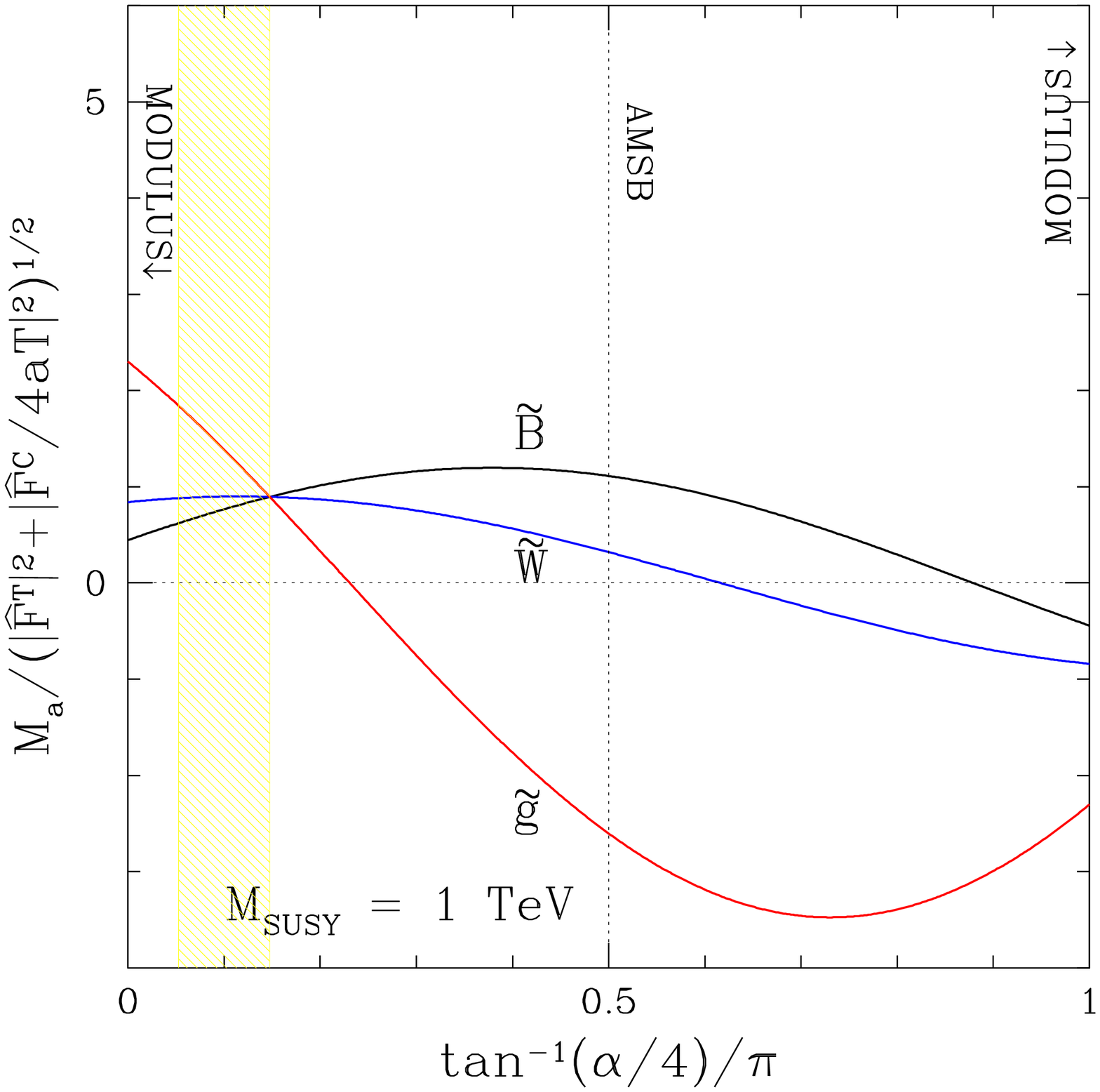,angle=0,width=8.0cm}}
{\hspace*{-.2cm}\psfig{figure=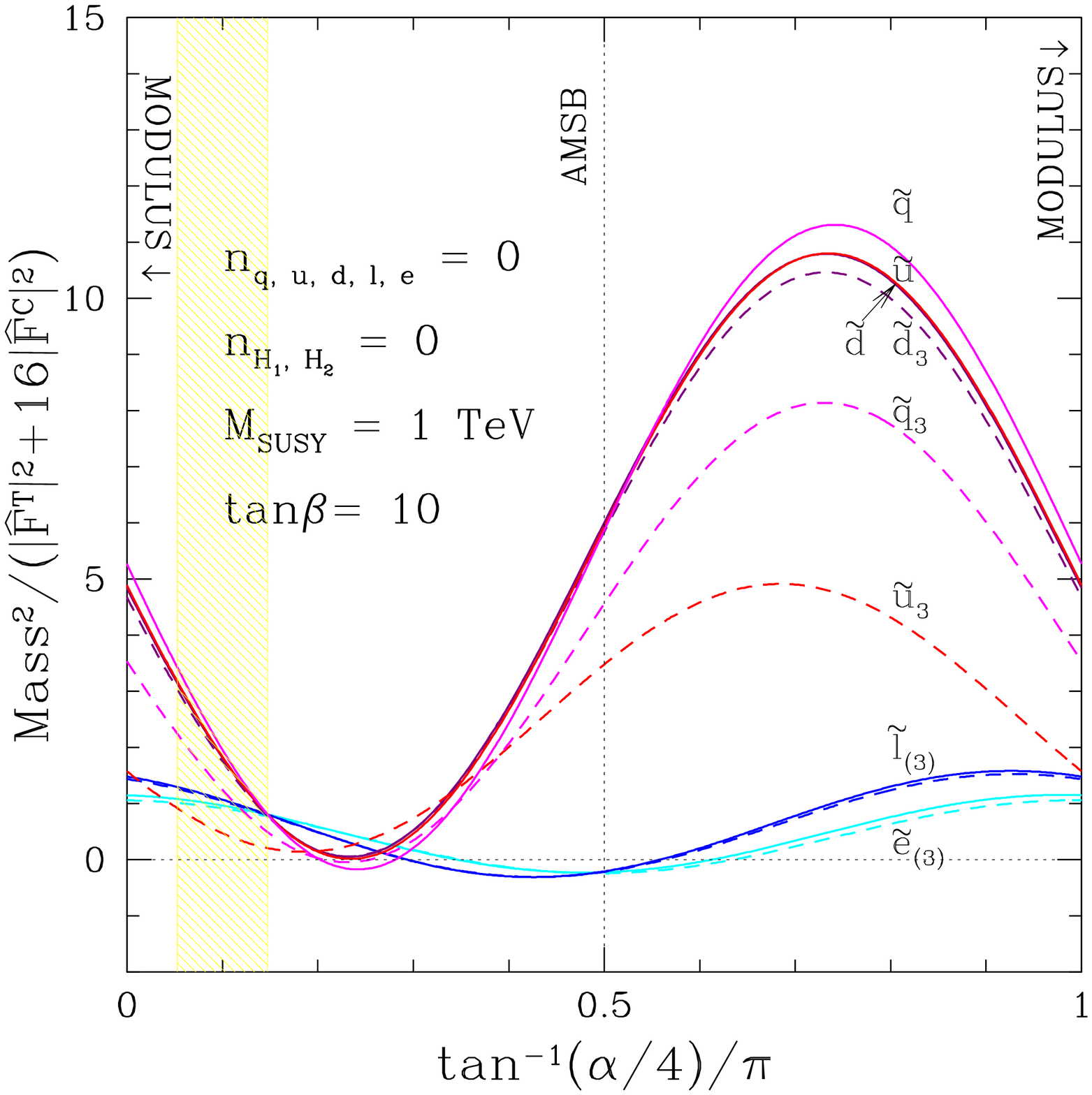,angle=0,width=8.0cm}}
} \caption{The pattern of superparticle masses at
$M_{SUSY}=1$ TeV for the entire range of the anomaly to moduli
mediation ratio $\alpha$.
Note that the sign convention of the gaugino masses  for $0\leq \tan(\alpha/4)\leq
\pi/2$ is different from the convention for
$\pi/2\leq \tan(\alpha/4)\leq \pi$. Here we choose $a_{IJK}=3$ and
$n_I\equiv 1-c_I=0$.
The shaded region indicates the range $2/3\leq \alpha\leq 2$ and
the short-dashed curves denote the 3rd generation
squarks/sleptons.
% Note that the sign convention of the gaugino
%masses (and $A_{ijk}$) for $0\leq \tan(\alpha/4)\leq \pi/2$ is
%different from the convention for $\pi/2\leq \tan(\alpha/4)\leq\pi$.
\label{fig:tree_rewsb}}
\end{minipage}
\end{center}
\end{figure}

\begin{figure}[t]
\begin{center}
\begin{minipage}{16cm}
\centerline{
{\hspace*{-.2cm}\psfig{figure=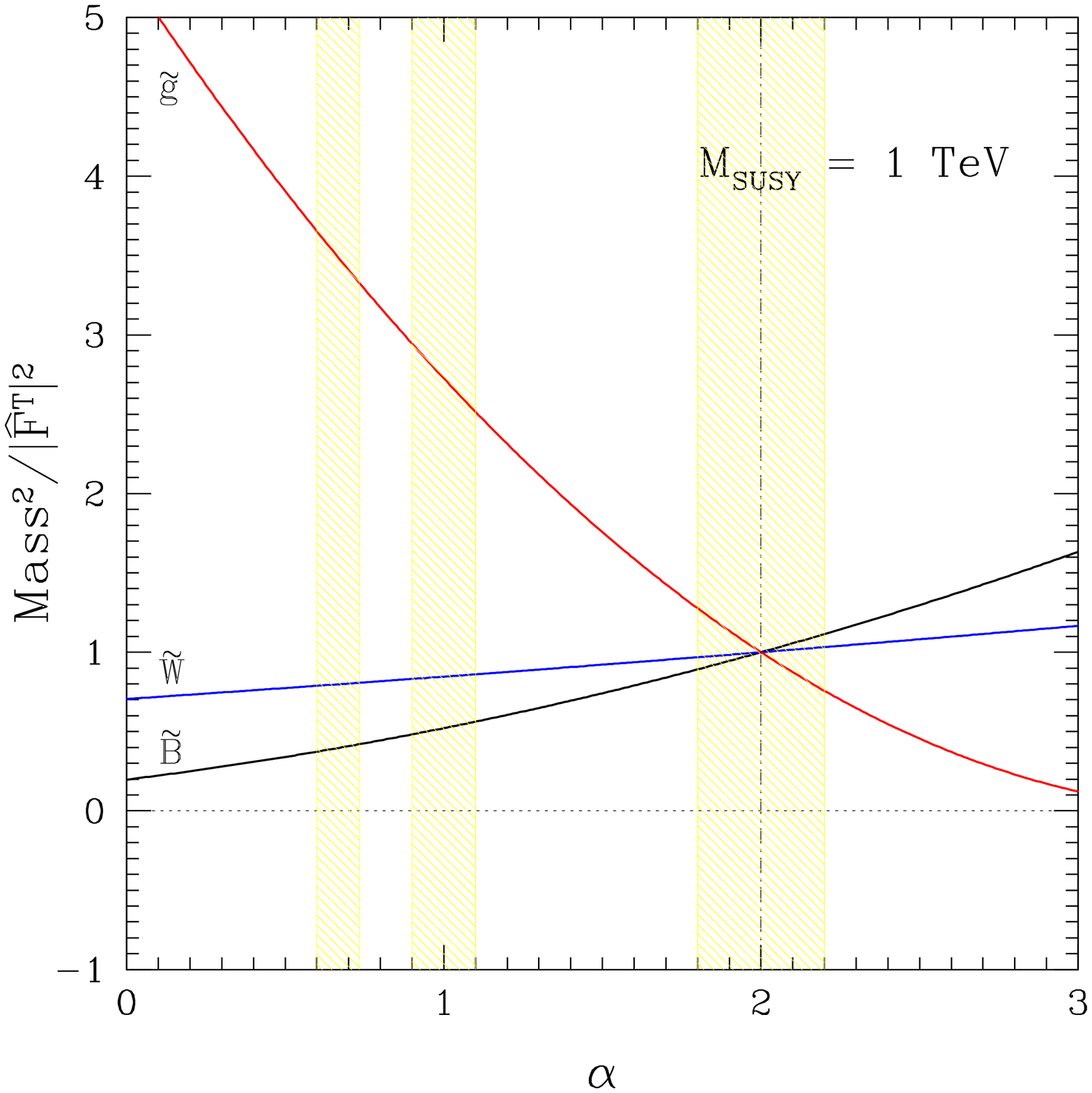,angle=0,width=8.0cm}}
{\hspace*{-.2cm}\psfig{figure=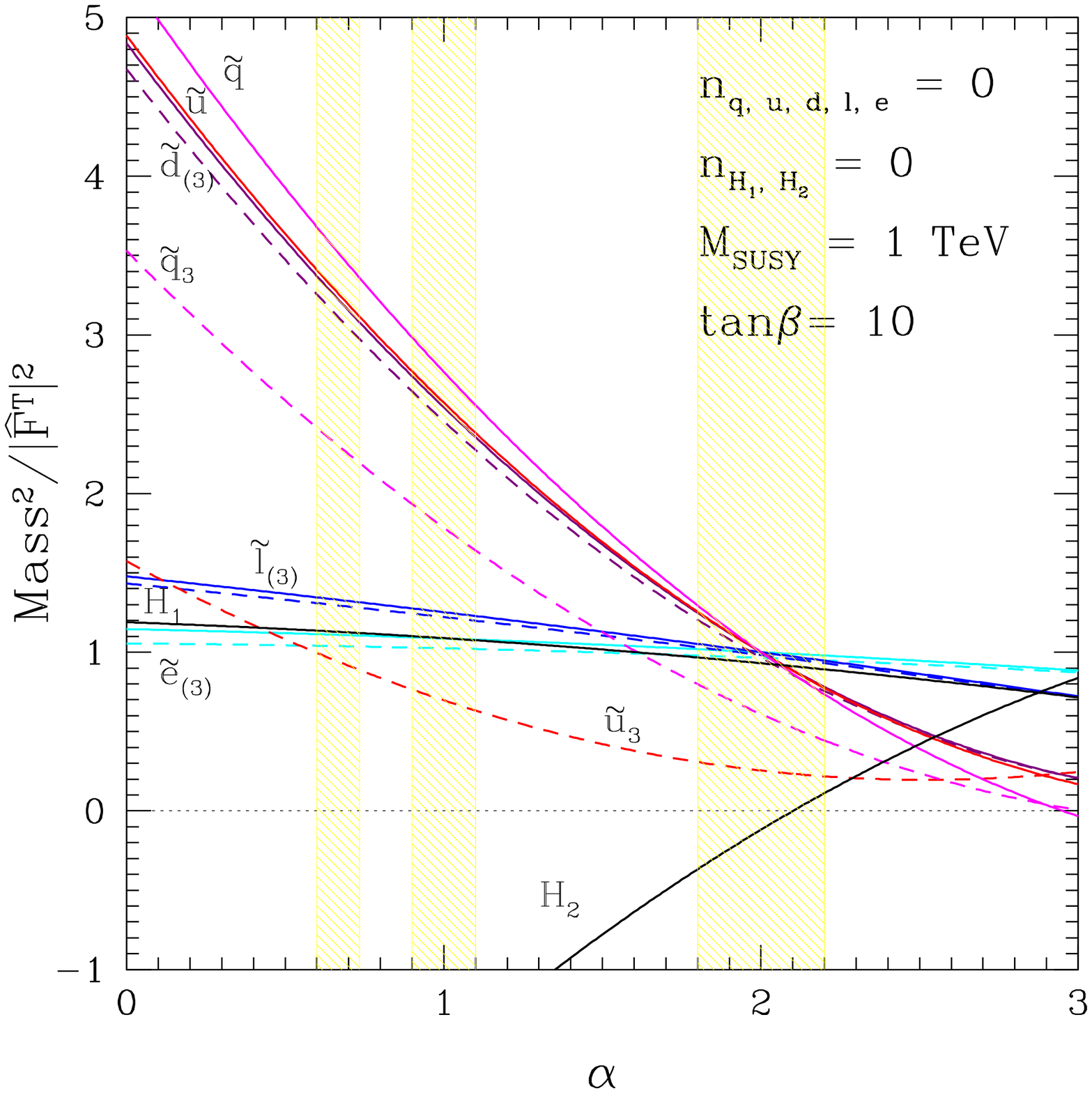,angle=0,width=8.0cm}}
}
%\caption{
%\label{fig:tree_rewsb}}
\end{minipage}
%\end{center}
%\end{figure}
%
%\begin{figure}[t]
%\begin{center}
%\begin{minipage}{16cm}
%\centerline{
%{\hspace*{-.2cm}\psfig{figure=d72c.eps,angle=0,width=8.0cm}}
%{\hspace*{-.2cm}\psfig{figure=d70c.eps,angle=0,width=8.0cm}} }
\caption{Superparticle masses at $M_{SUSY}=1$ TeV for $0\leq
\alpha\leq 3$. The shaded regions correspond to $\alpha=2/3, 1,
2$, taking into account $10\%$ uncertainty. Again the short-dashed
curves denote the 3rd generation squarks/sleptons. \label{fig:spectrum}}
%\end{minipage}
\end{center}
\end{figure}

In Fig.1, we depict how the pattern of low energy superparticle masses
varies as a function of $\alpha$.
Fig.2 shows the low energy superparticle  masses for $0\leq \alpha \leq 3$,
which contains the range of $\alpha$ suggested by the KKLT-type stabilization.
Particularly interesting values of $\alpha$ would be
$\alpha=1$ which is predicted by the KKLT-type
models  (\ref{kklt1}) and (\ref{kklt2}) with
$n_0=3$, $\ell=2$ and $f_a=T$,
and also $\alpha=2$ which gives nearly unified
gaugino (and sfermion) masses  at TeV scale.
Note that this pattern of low energy superparticle masses in
the mirage mediation are qualitatively different
from those predicted by other  SUSY breaking scenarios
such as the mSUGRA, gauge-mediation and anomaly-mediation.

We finally remark that the moduli stabilization schemes
discussed in section III can avoid dangerous SUSY flavor and CP violations
in a natural manner  even when all superparticle masses are
close to the weak scale \cite{choi1}.
The resulting (mixed moduli-anomaly mediated) soft terms  preserve the quark and lepton
flavors if the modular weights of the matter fields are flavor universal,
which would be achieved if the matter fields with common gauge charge
originate from the same geometric structure.
They also preserve CP
since the relative CP phases between $F^i$ and $F^C$ can be
rotated away by the shift of the axion-component of $T_i$
\cite{choi1,yama,choi3}.

\pagebreak
\vspace{5mm}
\noindent{\large\bf Acknowledgments}
\vspace{5mm}

This work is supported by the KRF Grant funded by
the Korean Government (KRF-2005-201-C00006),
the KOSEF Grant (KOSEF R01-2005-000-10404-0),
and the Center for High Energy Physics of Kyungpook National University.
I thank the organizers of the Summer Institute 2005 for the hospitality during the
workshop, and A. Falkowski, K. S. Jeong, L. Kallosh, T. Kobayashi, A. Linde,
H. P. Nilles, K. Okumura and
M. Yamaguchi for the collaborations and/or useful discussions.

\end{document}